# An HPC-Based Hydrothermal Finite Element Simulator for Modeling Underground Response to Community-Scale Geothermal Energy Production

Xiang Sun[1,4], Kenichi Soga[1], Alp Cinar[1], Zhenxiang Su[1], Kecheng Chen[1], Krishna Kumar[2], Patrick F. Dobson[3], and Peter S. Nico[3]

1. University of California, Berkeley, CA, USA
2. University of Texas, Austin, TX, USA
3. Lawrence Berkeley National Laboratory, Berkeley, CA, USA
4. Dalian University of Technology, Dalian, China

xs260@dlut.edu.cn

**ABSTRACT**

Geothermal heat, as renewable energy, shows great advantage with respect to its environmental impact due to its significantly lower $CO_2$ emissions than conventional fossil fuel. Open and closed loop geothermal heat pumps, which utilize shallow geothermal systems, are an efficient technology for cooling and heating buildings, especially in urban areas. An integrated use of geothermal energy technologies for district heating, cooling, and thermal energy storage can be applied to optimize the subsurface for communities to provide them with multiple sustainable energy and community resilience benefits. The utilization of the subsurface resources may lead to a variation in the underground environment, which might further impact local environmental conditions. However, very few simulators can handle such a highly complex set of coupled computations on a regional or city scale. We have developed a high-performance computing (HPC) based hydrothermal finite element (FE) simulator that can simulate the subsurface and its hydrothermal conditions at a scale of tens of km. The HPC simulator enables us to investigate the subsurface thermal and hydrologic response to the built underground environment (such as basements and subways) at the community scale. In this study, a coupled hydrothermal simulator is developed based on the open-source finite element library deal.II. The HPC simulator was validated by comparing the results of a benchmark case study against COMSOL Multiphysics, in which Aquifer Thermal Energy Storage (ATES) is modeled and a process of heat injection into ATES is simulated. The use of an energy pile system at the Treasure Island redevelopment site (San Francisco, CA, USA) was selected as a case study to demonstrate the HPC capability of the developed simulator. The simulator is capable of modeling multiple city-scale geothermal scenarios in a reasonable amount of time.

## 1. INTRODUCTION

Shallow geothermal energy (<400 m depth) has been adopted worldwide for heating and cooling buildings over the last decades with the application of ground source heat pump (GSHP) and groundwater heat pump (GWHP) systems (e.g., Freeston, 1996; Allen et al., 2003; Lund et al., 2011; Antics et al., 2016; Lund and Boyd, 2016; Sanner, 2016). In some countries, particularly in Austria, Germany, the UK and Switzerland, many new energy geotechnics technologies such as GSHP coupled deep and shallow foundations, diaphragm walls, tunnel linings and anchors have been applied in underground geotechnical construction (Brandl, 2006; Adam and Markiewicz, 2009; Bourne-Webb et al., 2009; Amatya et al., 2012; Laloui and Di Donna, 2013; Moormann et al., 2016; Soga and Rui, 2016; Mortada et al., 2018).

Recent studies have demonstrated that anthropogenic heat flow from buildings influences the subsurface temperature, especially in proximity of urban aquifers (Bidarmaghz et al., 2019a&b). Maps of geothermal potential may serve as a tool for the identification of geoenergy sources as well as for land development and spatial planning at both the regional and urban scales (Epting et al., 2013; Zhang et al., 2014; García-Gil et al., 2015). Although many researchers have proposed different methods to integrate the geological, hydrogeological and geophysical data into city-scale geothermal systems, few studies considered the actual ground temperature response or ground thermal condition variation resulting from the operation of geothermal systems such as thermal exploitation of a large number of installations (García-Gil et al., 2014; Zhang et al., 2015; Barla et al. 2020).

For any regional or city-scale geothermal installation, there are multiple different possible designs that will in turn couple in different ways with natural and anthropogenic subsurface variability to create a complex multiparameter space that must be explored for design optimization. In addition to natural subsurface variability, the uneven spatial distribution of excess heat in the subsurface can be viewed analogously to that of the urban heat island effect observed in many mega cities. With correct planning, this excess and unevenly distributed heat energy can be harnessed beneficially. However, if unchecked, it can result in high environmental and economic costs. If the current spatio-temporal variability of ground temperatures is known and future changes can be predicted, it would enable sustainable and resilient planning of underground activities in communities for both the short and long-term. To investigate such large city-scale underground thermal behavior, an HPC-based hydrothermal subsurface simulator is needed.

In this study, a coupled hydrothermal simulator based on the open source finite element library deal.II was developed to estimate the heat response of subsurface system on a city/regional scale. The performance of the developed code was verified using a benchmark case study and compared to that of the commercial software COMSOL Multiphysics. To test the capacity of the HPC code, the use of a large scale energy pile system at Treasure Island (San Francisco, CA, USA) was selected as a case study. The performance testing of this case study model with multiple processors show that multiple city-scale scenarios could be explored in a reasonable amount of time using this new code.

## 2. MODEL ESTABLISHMENT AND VERIFICATION

Subsurface heat transfer is affected by groundwater flow. When the groundwater flow velocity is high, the heat will transfer faster. In this study, it is assumed that the problem is a one-way coupled hydrothermal process. The governing equations of the process are shown in Eqs. 1 and 2.

$$\frac{1}{B_{poro}} \frac{dp}{dt} + \nabla \cdot \left(-k \cdot (\nabla p + \rho_w \mathbf{g})\right) = 0 \tag{1}$$

$$\frac{dT}{dt} + \nabla \cdot \left(-\frac{\lambda}{c_T} \nabla T\right) + \frac{c_w}{c_T} \mathbf{q} \cdot \nabla T = 0 \tag{2}$$

where

$$\lambda = (1-\phi)\lambda_s + \phi\lambda_w$$

$$c_T = (1-\phi)c_s + \phi c_w$$

$$k = \frac{K_i}{\mu_w} = \frac{K}{\rho_w g}$$

$$\mathbf{q} = -k(\nabla p + \rho_w \mathbf{g})$$

where p is the water pressure, T is the temperature, $\lambda_s$ is the thermal conductivity of solid, $\lambda_w$ is the thermal conductivity of pore fluid, $\phi$ is the porosity, $c_s$ is the heat capacity of solid, $c_w$ is the heat capacity of pore fluid, $B_{poro}$ is the porous compressibility, $\mathbf{g}$ is the gravity acceleration, $K_i$ is the permeability, $\mu_w$ is the viscosity of pore fluid, $\rho_w$ is the density of pore fluid, $K$ is the hydraulic conductivity and q is the Darcy's flow velocity.

To obtain a linear system for the subsurface heat and fluid transfer, the discrete form of the governing equations needs to be established. The finite element method with a backward Euler integral scheme is applied to obtain the discrete form of the governing equations given in Eqs. (1) and (2) as follows:

$$hkB_{poro} \cdot (\nabla \mathbf{N_p}, \nabla \mathbf{N_p})_\Omega \cdot \mathbf{P}^{n+1} + (\mathbf{N_p}, \mathbf{N_p})_\Omega \cdot \mathbf{P}^{n+1}$$
$$= -hB_{poro}(\mathbf{N_p}, \mathbf{q}^{n+1})_{\partial\Omega} + (\mathbf{N_p}, \mathbf{N_p}) \cdot \mathbf{P}^n - (\nabla \mathbf{N_p}, hkB_{poro} \cdot (\rho_w \mathbf{g}))_\Omega \tag{3}$$

$$(\mathbf{N_T}, \mathbf{N_T})_\Omega \cdot \mathbf{T}^{n+1} + \frac{h\lambda}{c_T}(\nabla \mathbf{N_T}, \nabla \mathbf{N_T})_\Omega \cdot \mathbf{T}^{n+1} + \left(\mathbf{N_T}, \frac{hc_w}{c_T}\left(-k\left(\nabla \mathbf{N_P} \cdot \mathbf{P}^{n+1} + \rho_w \mathbf{g}\right)\right) \cdot \nabla \mathbf{N_T} \cdot \mathbf{T}^{n+1}\right)$$
$$= -\frac{h}{c_T}(\mathbf{N_T}, -\lambda \nabla \mathbf{N_T})_{\partial\Omega} \cdot \mathbf{T}^{n+1} + (\mathbf{N_T}, \mathbf{N_T})_\Omega \cdot \mathbf{T}^n \tag{4}$$

where $\mathbf{N_p}$ and $\mathbf{N_T}$ are the weight functions for discretized pressure and temperature, respectively, n is the nth time step, $(*,*)_\Omega$ is the inner product operator and h is the time step size.

A high-performance computing system has thousands of cores and millions of elements, which can provide enough computing power to handle the proposed large-scale simulations. Many software products now have the feature of multi-threads and multi-nodes parallel computing compatibility such as COMSOL Multiphysics, Abaqus and Ansys-Fluent, etc. However, these commercial software products have low flexibility in the further development of the model. deal.II is the successor to the Differential Equations Analysis Library, which is a C++ programming library and is used to solve partial differential equations (PDEs) numerically (Arndt et al., 2020). The structure of

deal.II is flexible and it has a powerful computational resource optimization with popular open-source libraries on parallel computing. A preferable computing scheme can be selected to obtain the optimal solution for a specific problem of interest. Figure 1 shows the programming scheme of deal.II

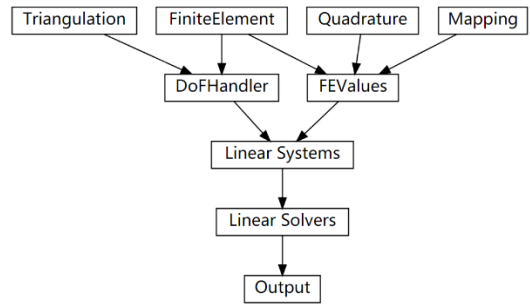

**Figure 1: The programming scheme of deal.II (Arndt et al., 2020)**

To verify the performance of the developed code, a benchmark case study is set up and the results of the developed code are compared to those obtained using the commercial FE analysis software COMSOL Multiphysics. The benchmark case simulates the variation of underground temperature and pressure during hot water injection. A large-scale three-layer model is built as shown in Figure 2. The size of the model is 20 m in radius and 300 m in depth. A 6-month hot water injection is simulated. Due to the axisymmetric nature, a small angle is set in the circumferential direction to reduce the scale of the model. There are three formation layers in the model; (i) caprock at the top, (ii) aquifer in the middle, and (iii) basement rock at the bottom. The model parameters are listed in Table 1. The hydraulic conductivity of the aquifer is set to $1 \times 10^{-6}$ m/s. For the caprock and basement rock, the hydraulic conductivity is set to $1 \times 10^{-10}$ m/s. and they are recognized as impermeable formations. Hot water of 288.15K is injected into the aquifer from the wellbore.

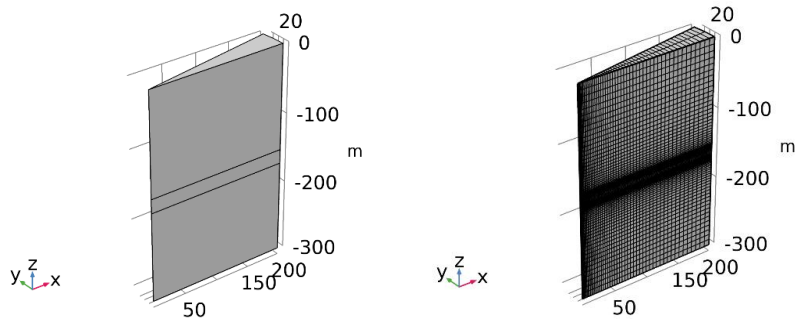

**Figure 2: Geometry and mesh of the model**

Sun et al.

**Table 1: Parameters of materials**

| Parameter | Value | Unit | Comment |
|---|---|---|---|
| $c_T$ | $1.2 \times 10^6$ | J/K/m$^3$ | Heat capacity for all formations |
| $\lambda$ | 1.2 | W/K/m | Heat conductivity for all formations |
| $B_{poro}$ | $1 \times 10^5$ | Pa | Porous compressibility |
| $T_{top}$ | 278.15 | K | Temperature at the top of model |
| $T_b$ | 288.15 | K | Wellbore temperature |
| $P_0$ | 0.1 | MPa | Initial pressure |
| $Q_b$ | 0.001 | m/s | Injection flux |
| $C_w$ | $1 \times 10^6$ | J/K/m$^3$ | Heat capacity of water |
| dT | 0.05 | K/m | Temperature gradient |

The simulation results of the temperature and pressure at 180 days are shown in Figure 3 and Figure 4, respectively. Heat flows into the aquifer with water from the wellbore and the temperature of the aquifer increases. Due to the low permeability of the caprock and basement rock, the heat in the vertical direction propagates slower than that in the radial direction. This is due to slow heat conduction in the vertical direction. The results given by deal.II and COMSOL yield similar trends. The temperature and pressure distributions along the centerline of the aquifer are given in Figure 5, whereas those along the wellbore in the vertical direction are shown in Figure 6. The comparisons show that the solutions given by deal.II match well with those computed by COMSOL. Table 2 shows the speed of computation using different codes on a PC and a node on HPC. The computation using our code based on deal.II is faster than COMSOL and the efficiency of parallelization in our code is better than COMSOL when running the same model on the same PC. The HPC node can provide much better computational power than the PC and the computation can be boosted by ten times faster when using single node on HPC.

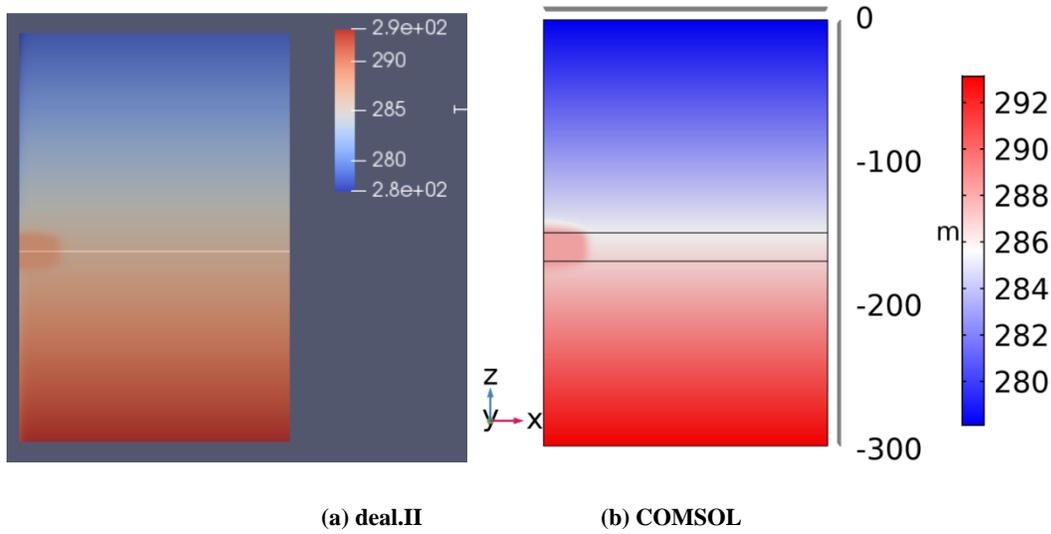

(a) deal.II  (b) COMSOL

**Figure 3: Comparison of temperature (K) between deal.II and COMSOL**

Sun et al.

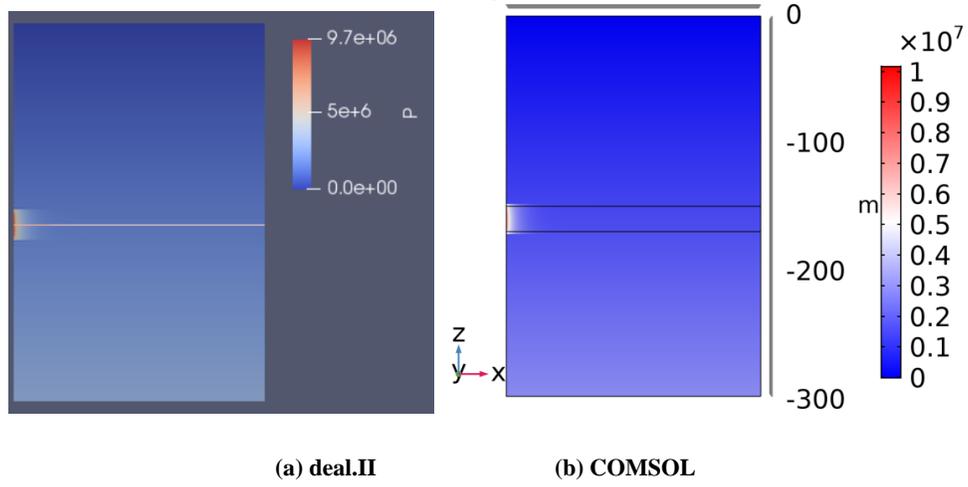

(a) deal.II  (b) COMSOL

Figure 4: Comparison of pressure (Pa) between deal.II and COMSOL

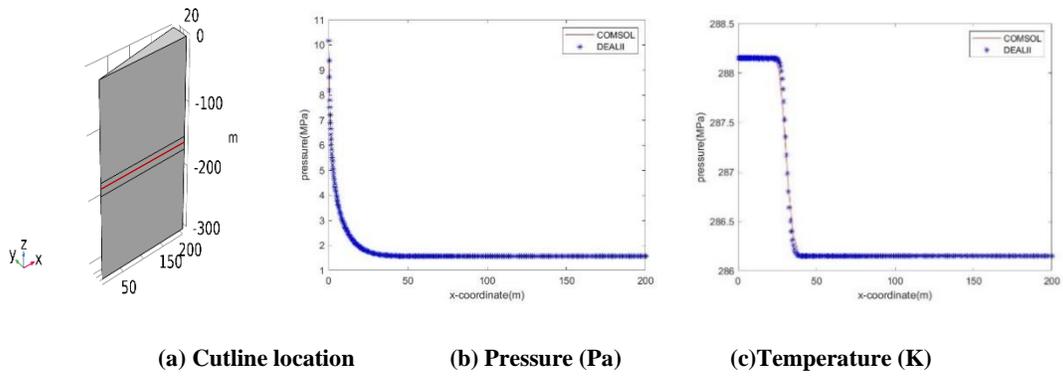

(a) Cutline location  (b) Pressure (Pa)  (c)Temperature (K)

Figure 5: Comparison of temperature and pressure distribution in the radial direction

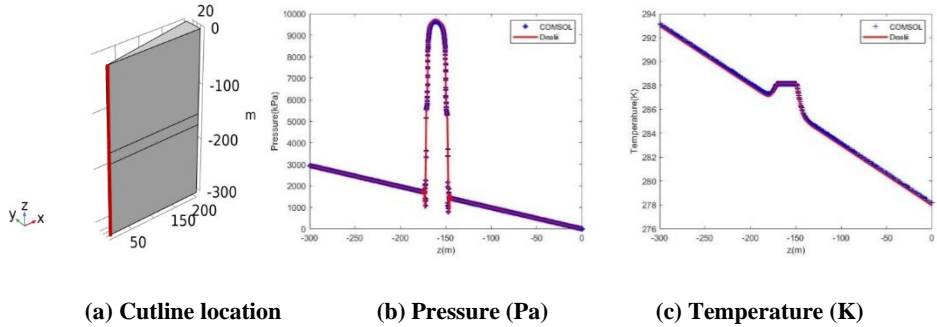

(a) Cutline location  (b) Pressure (Pa)  (c) Temperature (K)

Figure 6: Comparison of temperature and pressure distribution in the vertical direction

Table 2 Computation time spent on different computers with multiple processors

| Number of_MPI | COMSOL(PC) i7 8750H | Computation time (PC) i7 8750H | Computation time (HPC 1 node) |
|---|---|---|---|
| 1 | 1553 s | 423 s | 46.0 s |
| 2 | 1284 s | 269 s | 24.5 s |
| 4 | 1230 s | 237 s | 19.6 s |
| 6 | 1496 s | 240 s | 17.9 s |
| 8 | 1502 s | 262 s | 16.7 s |

Sun et al.

## 3. TREASURE ISLAND CASE STUDY

### 3.1 Introduction

Treasure Island is an artificial island that is a part of District 6 of the City and County of San Francisco. As shown in Figure 7, the proposed residential diversity of the Treasure Island site includes the integration of mid-rise buildings with high-rises. Townhomes and stack flats are also major components of this new development project. Because there are multiple types of buildings planned to cover a major portion of Treasure Island, their energy needs are also one of the concerns of the project developers. Application of state-of-the-art and clean energy production systems that have the capability of meeting the energy demands of the whole island community has been considered (Cinar et al., 2020). The usage of district-scale ground source heat pump system was considered as a possibility and this study considers this hypothetical implementation as a case study.

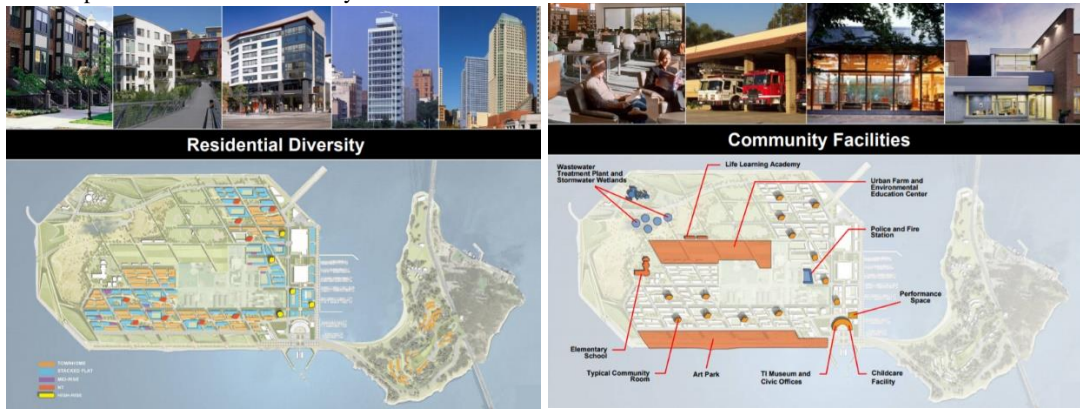

**Figure 7: The Treasure Island Redevelopment Project (ENGEO Inc 2009)**

### 3.2 Shallow Closed-loop Geothermal Energy Piles

Ground Source Heat Pumps (GSHPs) provide sustainable heating and cooling energy for housing, offices and retail spaces. For new building and underground developments, it is possible to incorporate the primary heat exchangers through the foundation elements (e.g., piles and basement walls) or into the tunnel linings. They are called energy (or thermal) piles/walls/tunnels.

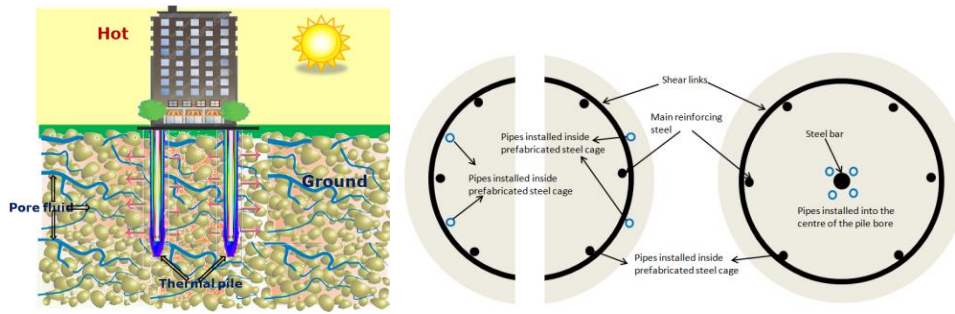

**Figure 8: Heat transfer of thermal pile (Soga and Rui, 2016)**

In an energy pile system, the heat exchanger pipe network is installed in the structural piles of a building as shown in Figure 8. Heat exchanger pipes are attached to the reinforcement cage of the pile before concrete is poured to create the pile, as shown in Figure 9. As concrete has excellent thermal conductivity and good heat storage properties, foundation piles are an ideal medium for heat transfer into the surrounding ground.

Sun et al.

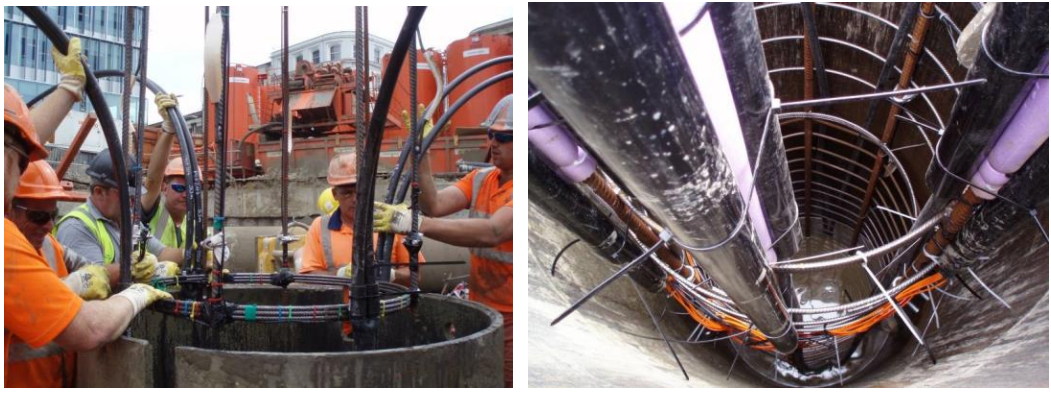

**Figure 9: Installation of geothermal loop into the energy pile (courtesy of Kenichi Soga)**

In this study, it is considered that a series of energy piles are installed at the footprint area where the new buildings are planned to be constructed as shown in Figure 7. A total of 1130 energy piles were integrated into the model and their locations are shown in Figure 10 (a).

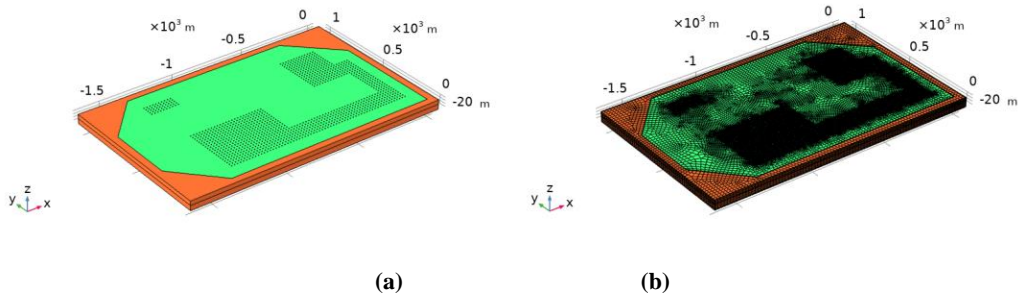

(a)                     (b)

**Figure 10: Treasure Island modeling (a) geometry model in which the dots are the locations where energy piles are installed and (b) mesh, which is dense around the piles and coarse around far-end boundary**

### 3.3 Finite Element Model

As shown in Figure 10(a), the length of the island in x-direction is 1680 m, the width of the island in y-direction is 1040 m, and the depth of the island in z-direction is 100 m. The whole island area is around 400 acres. From the edge of the island to the boundary of the whole model, a lateral margin of 45.7 m in width is extended to simulate the sea bed close to the island. Since the depth of the seabed is around 6.1 m, which is small relative to the model depth, this depth difference is not modeled. It is assumed that the part of the island submerged below the sea-level for a long period of time has the same temperature of 287.45 K as the sea. The temperature of the island above sea level is given as model input (constant or seasonal changes). A series of energy piles are set beneath the buildings as shown in Figure 10(a). The spacing between the piles is 20 m. The radius of the piles is 0.75 m and the length is 60 m. In total, there are 1 130 piles in the simulation model. Figure 10(b) shows a fine mesh around the piles and a coarse mesh far from the piles. The number of degrees of freedom (DOFs) is 1,301,718.

For the underground geology, starting from the ground, the hydraulic fill was placed during the initial construction of the island (Cinar et al., 2020). Beneath the hydraulically placed fill, a thin layer of sand-shoal deposits is observed. Young Bay mud lies beneath the sand-shoal deposits. Old Bay deposits, which are mainly composed of clay, are overlain by the Young Bay mud. At the bottom of the underground profile, there exists Franciscan rock formation. Depending on the location, the Franciscan rock formation is observed at a depth of 70-80 m beneath the ground surface throughout the island. Running the developed three-dimensional model requires the input of hydrothermal properties of the corresponding soil types. Thermal properties for the soil types observed in Treasure Island have not been extensively studied and hence values from the literature for this location were not found.

The hydrothermal properties gathered from the literature are summarized in Table 3.

Sun et al.

**Table 3: Parameters of the material of Treasure Island**

| Soil/Rock type | Hydraulic conductivity (m/s) | Heat Capacity (J/m³K) | Thermal conductivity (W/m·K) |
|---|---|---|---|
| Fill | $1.1 \times 10^{-5}$ * | $2.5 \times 10^{6}$ # | 2.2 $ |
| Shoal | $3.5 \times 10^{-5}$ ** | $2.5 \times 10^{6}$ # | 2.2 $ |
| Young bay mud | $1.7 \times 10^{-9}$ *** | $3.5 \times 10^{6}$ ## | 1.5 $ |
| Old bay mud | $1.7 \times 10^{-9}$ *** | $3.5 \times 10^{6}$ ## | 1.7 $ |
| Franciscan bedrock | $1 \times 10^{-6}$ **** | $2 \times 10^{6}$ ### | 2.97 $$ |

**\* Ashford and Rollins (2002), \*\* Phillips et al. (1993), \*\*\* Nguyen (2006), \*\*\*\* Williamson (1991), # Russo et al. (2009), ## Goto et al. (2017), ### Konakova et al. (2013), $ Misra et al. (1995), $$ Walters and Combs (1991)**

The seasonal air temperature variation of Treasure Island is shown in Figure 11. It is adopted from the U.S. National Weather Service and the weather data is reported for San Francisco International Airport between 1985-2015. The average temperature in the whole year is 287.45 K. Due to the lack of ground surface temperature, the sunlight strength and local humidity records for the island, the air temperature record is assigned as the ground surface temperature.

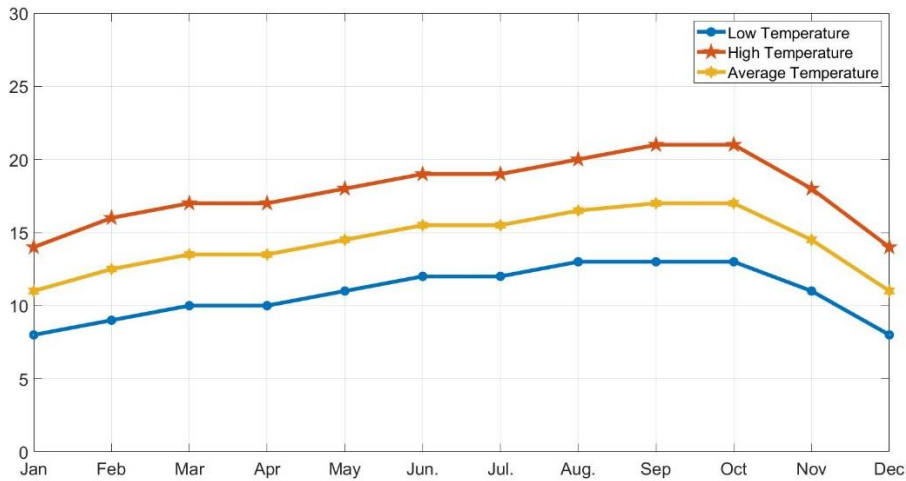

**Figure 11: Air temperature variation of San Francisco International Airport (from the U.S. National Weather Service), which is used as a proxy for the ground surface temperature of Treasure Island**

A relatively realistic surface temperature variation, which is an average temperature monthly in Figure 11, is applied on the top surface of the model. For the convenience of the simulation, the temperature record in May, which is equal to 287.65K and approximates the average temperature yearly, is chosen as the beginning of the simulation. The temperature gradient Tdrop is assumed to be 0.03K/m by assigning a constant flux at the bottom boundary. A heat flux qT of 0.89W/m², which is equal to the thermal conductivity times the temperature gradient around the bottom bedrock is applied to produce this temperature gradient. For the initial groundwater pressure distribution, the pressure values on the left-side boundary are assigned as hydrostatic from the ground surface, assuming the water table is at the top boundary surface. On the right-side boundary, the pressure value is assigned to be the sum of 8.5 kPa and the hydrostatic pressure, giving a hydraulic gradient Pdrop of 5 Pa/m from the right boundary to the left boundary. The boundary conditions are consistent with the initial conditions, which are shown in Figure 12. The heat flux from the bottom of the model remains as 0.89W/m². Meanwhile, the initial pile temperature is set to the average value of the initial underground temperature distributed along the pile, which equals 288.55K. The variation of the pile temperature is assumed to be a square wave of which the magnitude equals 13 K. This corresponds to a constant value of 301.55K for 6 months during summer and then a constant value of 275.55K for 6 months during winter.

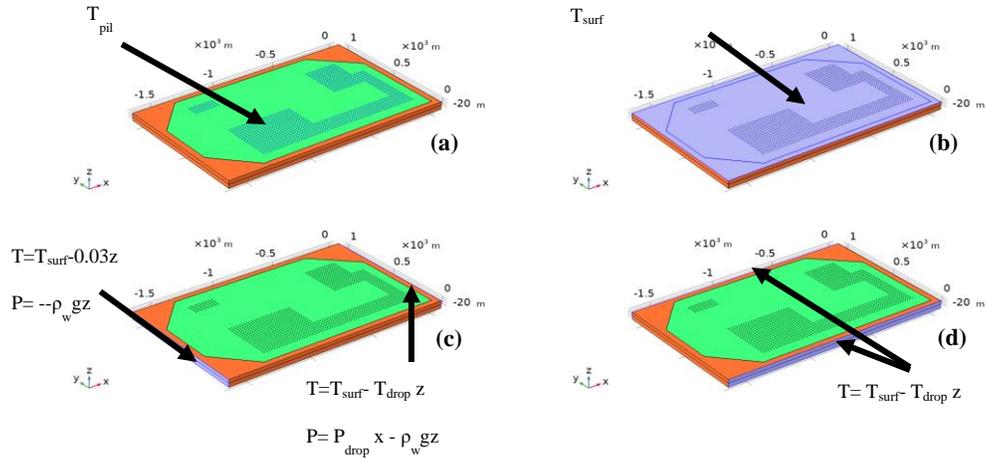

**Figure 12: Boundary conditions (a) at the piles; (b) on the top of the model; (c) on the right and left side of the model; (d) on the front and back of the model**

### 3.4 Parallelization

To simulate the heat transfer in such a large-scale underground model, a parallel computing scheme needs to be implemented. There are two ways to utilize multi-processor machines to carry out the parallelization: (1) Shared memory parallelization in which each machine keeps the entire mesh and DoF handler locally, but only a share of the global matrix, sparsity pattern, and solution vector is stored on each machine and (2) Distributed memory parallelization in which the mesh and DoF handler are distributed, i.e., each processor stores only a share of the cells and degrees of freedom. No processor knows the entire mesh, matrix, or solution. Problems solved in this mode are usually large (100s of millions to billions of degrees of freedom) that no processor can or should store even a single solution vector.

Here the distributed parallelization that is more efficient for the large number of DOFs problem is chosen. The scheme is shown in Figure 13. Each processor holds one part of the grid and deals with its own part. It reduces the usage of memory and the processor can save significant time on assembling the system because it just needs to traverse the grid passed into itself. Each processor processes its own part of the model and gets the solution by solving the linear system. All the solutions given by different processors are then combined together and outputted through one processor.

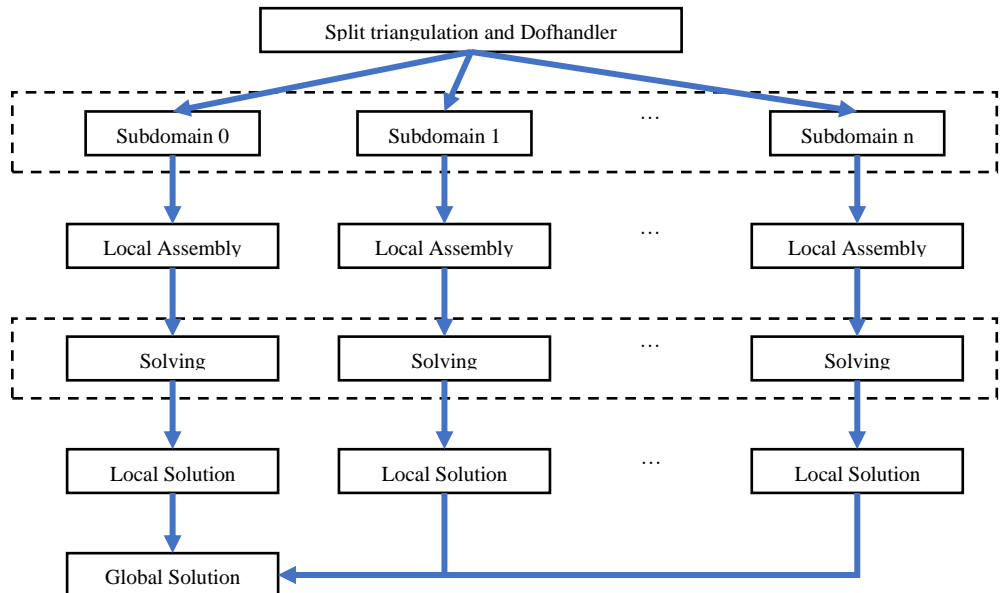

**Figure 13: Distributed parallelization scheme**

Sun et al.

## 3.5 Results

Figure 14 shows the variation of underground temperatures at different times of a year. Due to slow groundwater flow, the heat migration in the ground is diffusion dominated. During the first 180 days, the underground temperature is increasing with heat diffusion from the piles. After 180 days, the cooling phase begins and the ground temperature around the piles decreases and the cooling fronts propagate into the soil.

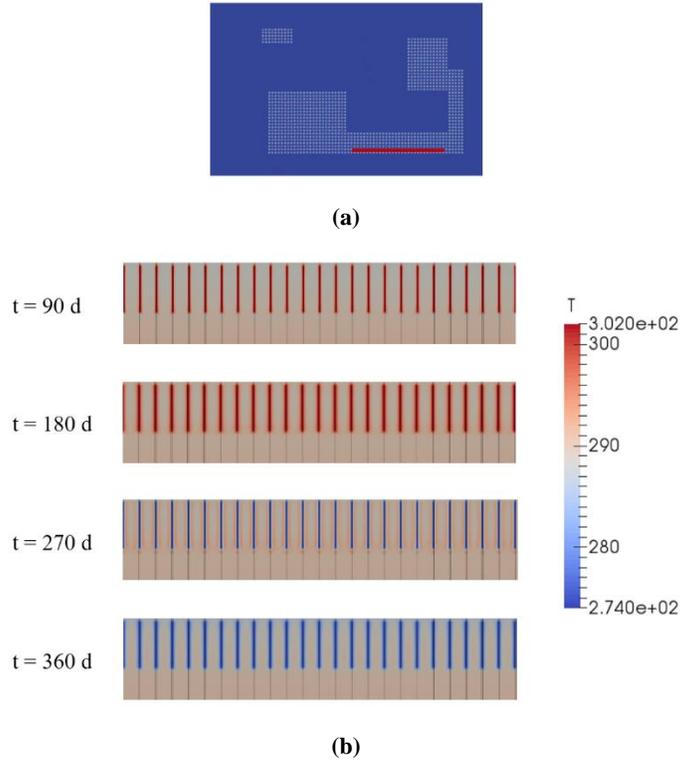

**Figure 14: Underground temperature variation (b) during the cyclic heating and cooling at the cross-section marked as red line on the island map (a) – CASE 3**

The observation lines shown in Figure 15 are selected to investigate the details of ground response to thermal loading by both the seasonal ground surface temperature variation and the heating and cooling of the energy piles. The observation lines O-1, O-2 and O-3 are set at the locations of 1 m, 5 m and 10 m away from the left pile, respectively. Because the underground temperature response profile is symmetric with respect to the center at the distance between two piles, it is not necessary to plot the right side of the center line Q-3. To investigate the temperature distribution in the horizontal direction, the observation lines Q-4, Q-5 and Q-6 are set at the location where the temperature responses are representative. For example, Q-4 is at 6 m depth from the surface because the temperature response here is impacted by both ground surface temperature variation and pile heating. Q-5 is at 40 m depth from the surface because this place is at the middle of where the underground temperature is impacted by the pile heating and cooling. Q-6 is at 60 m depth and is at the bottom of the pile with heat propagation into the bedrock.

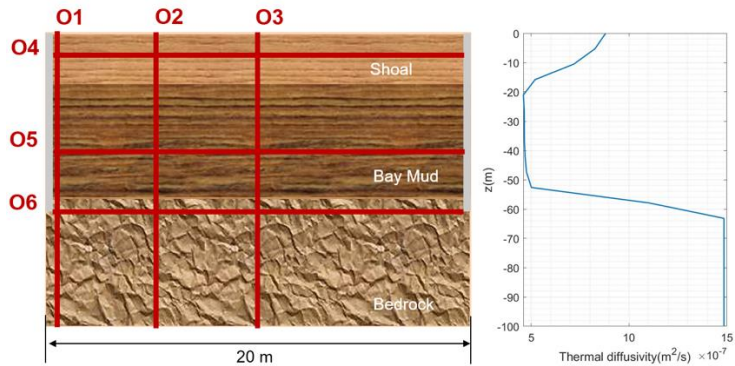

**Figure 15: Heat diffusivities corresponding to different layers and the observation locations in CASE-3**

Sun et al.

Figure 16 shows the temperature distributions after 90, 180, 270 and 360 days of heating (for 6 months) and then cooling (for 6 months) at the observation lines O-1, O-2 and O-3. Since heat conduction dominates the underground temperature response, the distribution of underground heat diffusivity plotted in Figure 15 is the key factor of impacting the underground temperature profile. The variation of temperature at shallow depths is sensitive to the surface temperature change as shown by comparing Figure 15 and Figure 16. However, the seasonal variation only affects the highly heat conductive layers including the filled and shoal formations. It has little impact on the low heat conductive clay layers. The clay layers are mainly affected by the heat injection from the energy piles.

Due to the low heat diffusivity, the injected heat takes a long time to reach to the observation location far from the piles as shown in Figure 16(b) and Figure 16(c). The underground temperature response at the vicinity of the piles is sensitive to the heat injection as shown in Figure 16(a). With the increase in the distance from the piles, the impact of heating and cooling on the underground temperature response becomes smaller. This is also observed in Figure 17. Although the pile temperature is decreasing, the temperature at the observation location Q-2 can be increasing due to the residual heating. The temperature at the observation location Q-3 has little change as it is far away from the piles, illustrating that the interaction of thermal loadings from the two adjacent energy piles is limited in this particular case. The temperature of the bed rock at the depth of 60 m changes faster than the clay layers with heating and cooling due to the high heat diffusivity of the bedrock as shown by comparing Figure 17(b) and Figure 17(c).

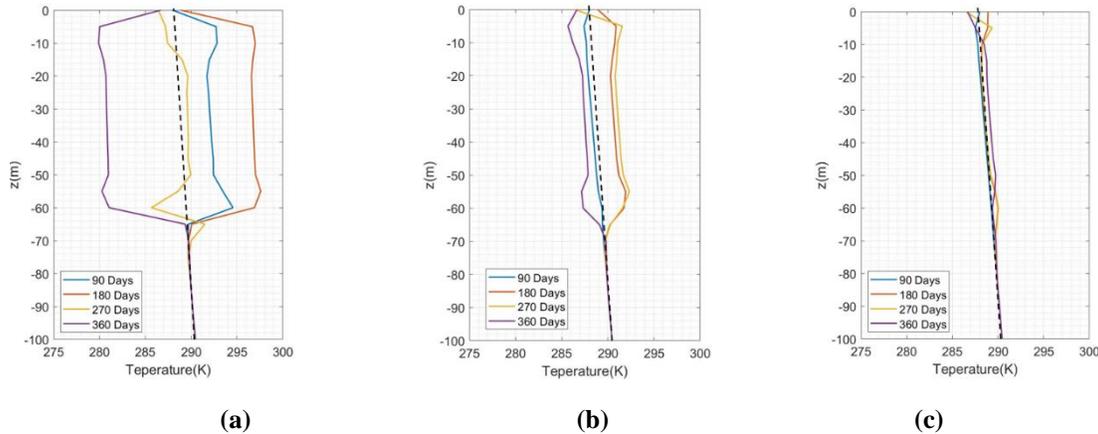

(a)　　　　　　　　　　　　　　　(b)　　　　　　　　　　　　　　　(c)

**Figure 16: Vertical subsurface temperature change (in z direction) with cyclic heat injection at the observation locations (a) O-1: 1 m, (b) O-2: 5 m and (c) O-3: 10 m far away from the left pile, respectively. The dashed back line is the initial temperature distribution at 0 days.**

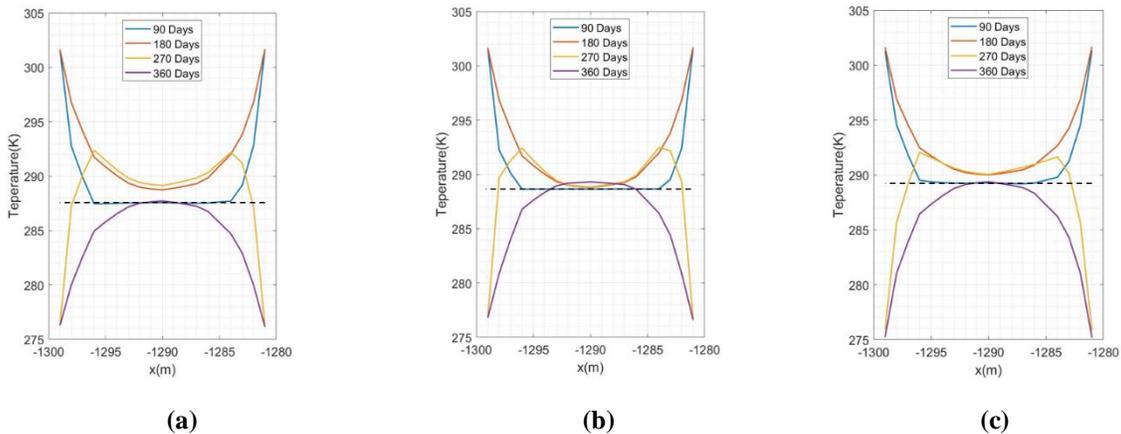

(a)　　　　　　　　　　　　　　　(b)　　　　　　　　　　　　　　　(c)

**Figure 17: Lateral subsurface temperature variation between two piles (in x direction) with cyclic heat injection at the observation locations (a) O-4: 6 meters, (b) O-5: 40 meters and (c) O-6: 60 meters depths, respectively. The dashed back line is the initial temperature distribution at 0 days**

The computational efficiency of the code for this Treasure Island problem is examined. Figure 18 (a) shows the computational time taken by different computing processes including assembling T system, assembling P system, solving T system and solving P system. Most of the computational time is taken to assemble the linear system. The total time spent on the Treasure Island simulation case decreases linearly with the number of the processors in the double logarithm scale as shown in Figure 18(b). However, with the increase in the number of the processors, the efficiency increases proportionally in the double log scale. For example, using 12 processors, it took 703

Sun et al.

seconds to solve a problem with a number of DOFs of 1,301,718. For the smaller problem (283,810 DOFs) shown in Figure 2, most of the computing time was spent on solving the heat equation. However, in the Treasure Island case, the time spent on solving the heat equation was less than the other processes. This illustrates that, as the model scale increases, assembling the matrix becomes the main time-consuming process due to traversing a large number of cells in the model. Results of this study show that, given the relatively short computing times using multiple processors, multiple scenarios can be explored for the Treasure Island case in a reasonable amount of time.

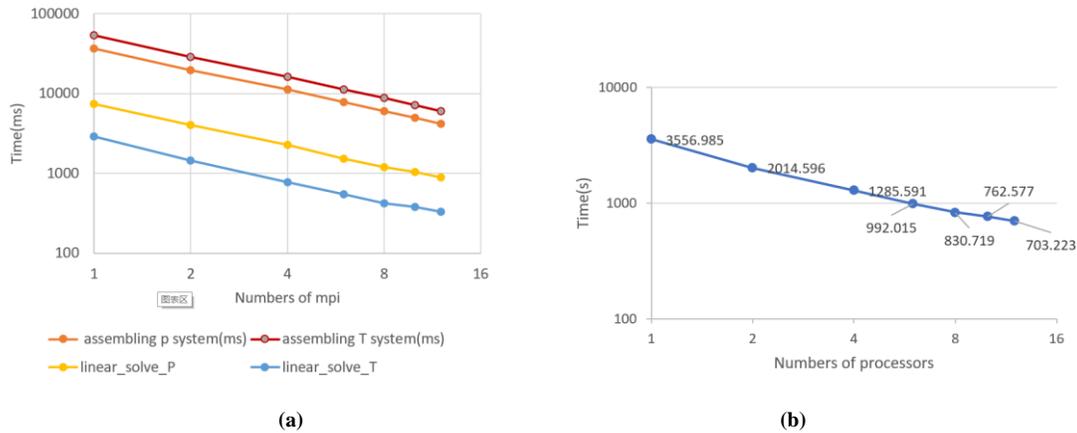

**Figure 18: The effect of the number of processors on computational time for the Treasure Island problem**

## 4. CONCLUSIONS

A high-performance computing based hydrothermal finite element simulator was developed so that a problem of city-scale geothermal utilization for community energy resilience can be investigated in the next stage of this project. The code uses an integrated parallelization scheme to compute the subsurface temperature and pressure variation during heat injection or extraction into the ground. The code leverages the deal.II finite element open-source library and the parallel programming library PETSc. The performance of the code was validated and tested by comparing the results to those of the commercial finite element software COMSOL Multiphysics. The improved computational efficiency of the new code with increasing processor number was demonstrated when compared to the efficiency of COMSOL. A city-scale Treasure Island model, which included 1130 energy piles installed in the island as part of its redevelopment project, was simulated to test whether the code can handle a large-scale simulation.

The preliminary validation of the code shows its promise but more testing is required to improve its computational efficiency. The assumptions made for the Treasure Island case are rather simplistic. The case for the Treasure Island is just the tip of the iceberg in terms of exploring variability at the moment, but because we were able to do the case 'relatively easily', this provides an initial test of the performance of the developed code. Future extension includes seasonal loading effects, the heterogeneity of model parameters, more realistic geometry, tidal effects, etc. Such work will be carried out in the next phase of the code development.

## ACKNOWLEDGMENTS

This work was supported by (i) the U.S. Department of Energy, Office of Energy Efficiency and Renewable Energy (EERE), Office of Technology Development, Geothermal Technologies Office (GTO), under Award Number DE-AC02-05CH11231 with LBNL and (ii) the National Science Foundation, Award Number #1903296 "CMMI-EPSRC: Modeling and Monitoring of Urban Underground Climate Change (MUC2)". We thank ENGEO for providing geotechnical details relating to the Treasure Island site.

Sun et al.